# Giant coercivity, resistivity upturn, and anomalous Hall effect in ferrimagnetic FeTb

Lijun Zhu,[*1,2] Lujun Zhu,[3] Qianbiao Liu,[1] Xin Lin[1,2]

1. State Key Laboratory of Superlattices and Microstructures, Institute of Semiconductors, Chinese Academy of Sciences, Beijing 100083, China
2. College of Materials Science and Opto-Electronic Technology, University of Chinese Academy of Sciences, Beijing 100049, China
3. College of Physics and Information Technology, Shaanxi Normal University, Xi'an 710062, China

Email: [*ljzhu@semi.ac.cn](*ljzhu@semi.ac.cn)

**Abstract:** Despite the blooming interest, the transition-metal rare-earth ferrimagnets have not been comprehensively understood in terms of their coercivity and transport properties. Here, we report a systematic study of the magnetic and transport properties of ferrimagnetic FeTb alloy by varying the layer thickness and temperature. The FeTb is tuned from the Tb- dominated regime to the Fe-dominated regime via the layer thickness, without varying the composition. The coercivity closely follows the $1/\cos\theta_H$ scaling (where $\theta_H$ is the polar angle of the external magnetic field) and increases quasi-exponentially upon cooling (exceeding 90 kOe at low temperatures), revealing that the nature of the coercivity is the thermally-assisted domain wall depinning field. The resistivity exhibits a quasi-linear upturn upon cooling possibly due to thermal vibrations of the structure factor of the amorphous alloy. The existing scaling laws of the anomalous Hall effect in the literature break down for the amorphous FeTb that are either Fe- or Tb-dominated. These findings should advance the understanding of the transition-metal-rare-earth ferrimagnets and the associated ferrimagnetic phenomena in spintronics.

## I. Introduction

Ferrimagnetic materials (FIMs), which have two antiferromagnetically coupled sublattices, are of considerable interest in the field of spintronics [1-5]. FIMs are potentially advantageous for dense magnetic recording applications [6] because of their tunable magnetism, less sensitivity to stray magnetic fields than ferromagnets (FMs), and easier and fast detection than antiferromagnets (AFs). Moreover, FIMs are an exotic platform to study the interplay of spin-orbit physics and AF coupling as a function of the degree of magnetic compensation. Several striking spin-orbit-coupling (SOC) phenomena have been demonstrated to arise from the magnetization compensation within the bulk of the FIMs, such as large magnetic domain wall velocities near compensation [3-5], strong compensation-dependence and sign reversal of bulk spin-orbit torques (SOTs)[7], strong variation of the "interfacial" SOTs with the relative spin relaxation rates within the bulk of FIMs [8,9], lack of current-driven magnetization switching at full magnetization compensation [10-12]. However, the spin-mixing conductance of the interfaces of metallic FIMs is insensitive to temperature and magnetic compensation or the areal density of the magnetic moment of the interface [8], in contrast to the insulating FIM case [13-14].

An in-depth understanding of the ferrimagnetic phenomena in magnetic heterostructures requires insights into the mechanisms of the coercivity ($H_c$), the electron momentum scattering, and the anomalous Hall effect (AHE) of the FIMs. Note that the coercivity of a perpendicular magnetization represents the switching barrier to overcome by the driving magnetic field or SOT [15-17], while electron momentum scattering affects the generation [18-20] and relaxation of spin current via SOC [8,21,22]. The AHE typically functions as the indicator of the magnetization orientation in a variety of experiments (e.g., harmonic Hall voltages [7-9] and magnetization switching [14-15,19,23,24]). However, the magnetization of transition-metal rare-earth FIMs arises from two competing sublattices that have been suggested to contribute to the AHE in distinct manners (i.e., the 3$d$ states of the transition metal governed the transport properties, while the 4$f$ states of Tb were less involved [23,25,26]). The scaling behavior of the AHE in such transition-metal rare-earth FIMs thus becomes stimulating open questions.

In this work, we systematically examine the magnetic and transport properties of FeTb, a representative FIM, as a function of layer thickness and temperature. We show that the coercivity can be rather high and increase quasi-exponentially upon cooling due to the nature of the thermally-assisted domain wall depinning field. The large longitudinal resistivity ($\rho_{xx}$) increases quasi-linearly upon cooling. The anomalous Hall resistivity ($\rho_{AH}$) cannot be described by the existing scaling laws [27-29].

## II. Samples and magnetization

For this work, we deposit four FeTb films with different thicknesses ($t$ = 8 nm, 16 nm, 32 nm, and 48 nm) on Si/SiO$_2$ substrates by co-sputtering at room temperature. Each sample is capped by a MgO (2 nm)/Ta (2 nm) bilayer that is fully oxidized upon exposure to the atmosphere. The composition of all the FeTb layers is Fe$_{0.55}$Tb$_{0.45}$ in volume percentage as calibrated using the deposition rates of the Fe and the Tb. This composition corresponds to a Tb/Fe atomic ratio of ≈ 0.3 (as calculated using the atomic volumes of bulk Fe and Tb crystals), which is consistent with the energy dispersive spectroscopy (EDS) results (0.32±0.01) in Figs. 1(a) and 1(b). Such FeTb films are amorphous and homogeneous as indicated by scanning transmission electron microscopy and electron energy loss spectra results of the samples prepared by the same sputtering tool using similar parameters within the same few days [7].

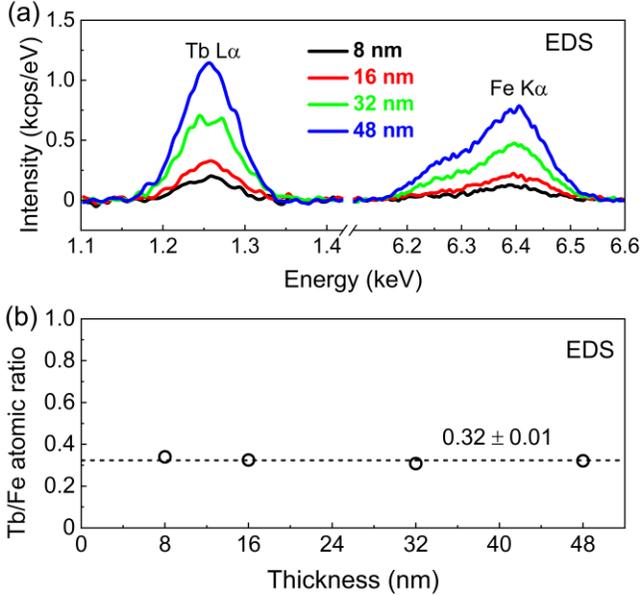

Fig. 1. (a) Energy dispersive spectra collected using a Bruker EDS tool and (b) the Tb/Fe atomic ratio calculated using the Tb L$\alpha$ and Fe K$\alpha$ EDS spectra for the FeTb films with different thicknesses, revealing that the composition is the same for the FeTb films studied in this work.

The saturation magnetization ($M_s$) for each FeTb film, the sum contribution of the Fe and the Tb sublattices (Fig. 2(a)), is measured using a superconducting quantum interference device (SQUID). Figure 2(b) shows the temperature profile of the saturation magnetization of the FeTb films with different thicknesses. $M_s$ for the 8 nm FeTb varies non-monotonically with temperature and peaks at 250 K. For the 16 nm FeTb, $M_s$ increases monotonically from being negligibly small at temperatures below 50 K to $\approx$60 emu/cm$^3$ as the temperature approaches 300 K. In contrast, for both the 32 nm and 48 nm FeTb, $M_s$ decreases first slowly and then more rapidly as the temperature increases. The diverse temperature profiles of the saturation magnetization of the FeTb films imply a strong tuning of the magnetization compensation configuration by thickness. As plotted in Fig. 2(b), the magnetization exhibits full compensation at the "compensation thickness" of 16 nm at 5 K and of $\approx$20 nm at 300 K. Thus, the magnetization compensation of the FeTb alloy is not only a function of substrate [7], composition [7,30,31], and strain [32] but also of temperature and thickness.

The films are then patterned into 5$\times$60 $\mu$m$^2$ Hall bars for electrical measurements using a physical properties measurement system (PPMS-9T). From the AHE measurements (see Figs. 2(c) and 2(d)), the FeTb films have fairly square hysteresis loops for the transverse resistivity ($\rho_{xy}$), implying good magnetic uniformity of these films. The polarity of the hysteresis loop is opposite for the 8 nm and 16 nm FeTb samples compared to that for the 32 nm and 48 nm ones, suggesting that the films are Tb-dominated at 8 nm and 16 nm but Fe-dominated at 32 nm and 48 nm. Such striking thickness dependences of the magnetization and the transverse resistivity are interesting observations and worth future investigation. While the unambiguous identification of the exact mechanism is beyond the scope of this paper, we speculate that the striking thickness dependence of the magnetic and transport properties might be related to some hidden short-range ordering within the amorphous films.

### III. Giant, strongly temperature-dependent coercivity

Figure 3(a) shows the out-of-plane coercivity and the effective perpendicular magnetic anisotropy field ($H_k$) at 300 K. Here, the out-of-plane coercivity is estimated from the switching of the transverse resistivity by a perpendicular magnetic field ($H_z$, Fig. 2(d)), while $H_k$ is estimated from the parabolic scaling of $\rho_{xy}$ with the in-plane magnetic field ($H_{xy}$) due to tilting of the magnetization (Fig. 3(b)), i.e.,

$$\rho_{xy} = \rho_{AH} \cos(\arcsin(H_{xy}/H_k)) \approx \rho_{AH} [1-1/2(H_{xy}/H_k)^2]. \quad (1)$$

Both $H_c$ and $H_k$ vary as a function of the layer thickness and tend to increase upon approaching the "compensation thickness". As shown in Fig. 3(c) and 3(d), $H_c$ follows a 1/cos$\theta_H$ scaling ($\theta_H$ is the polar angle of the driving magnetic field $H_{xz}$) and is typically much smaller than the perpendicular magnetic anisotropy field at $\theta_H = 0°$. This is in contrast to the case of a perpendicular macrospin, for which the coercivity varies as

$$H_c = H_k (\cos^{2/3}\theta_H + \sin^{2/3}\theta_H)^{-3/2}, \quad (2)$$

and is equal to $H_k$ at $\theta_H = 0°$ and 90°. As plotted in Fig. 3(e), the out-of-plane coercivity of each FeTb film increases upon cooling in a quasi-exponential manner. These observations consistently reveal that the coercivity of the FeTb represents the thermally-assisted depinning field of the magnetic domain walls rather than the bulk magnetic anisotropy field. This is generally the case for magnetic systems in which the reversed domain nucleation and domain wall propagation (with the energy barrier of the domain wall pinning field) require less energy than coherent rotation (with the energy barrier of $H_k$) [17].

We also note that the out-of-plane field required to switch these films, i.e., the coercivity, exceeds 90 kOe at temperatures below 175 K for the 16 nm FeTb and at temperatures below 25 K for the other three samples. The rapid increase of the coercivity upon cooling is distinct from the enhancement of coercivity at the magnetization compensation points [1-5,7, 33] because it occurs in the whole temperature region, including the temperatures at which the magnetization is very high (Fig. 2(b)). The giant coercivity and square hysteresis loops (Fig. 3(f)) may make these FeTb interesting hard magnets for some specific spintronic applications [34].

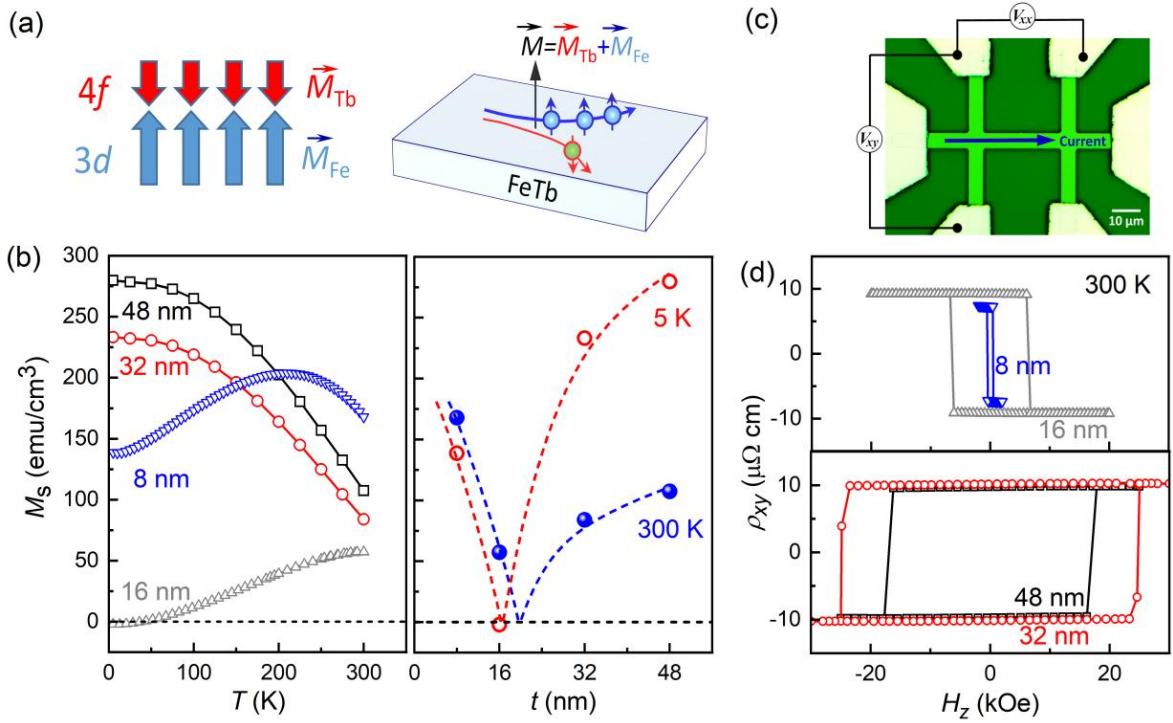

Fig. 2. (a) Schematic depict of the ferrimagnetism and the anomalous Hall effect in the FeTb. (b) Dependence of the saturation magnetization on temperature, (c) Dependence of the saturation magnetization on the thickness at 5 K and 300 K, and (d) Transverse resistivity at 300 K for the FeTb with different layer thicknesses.

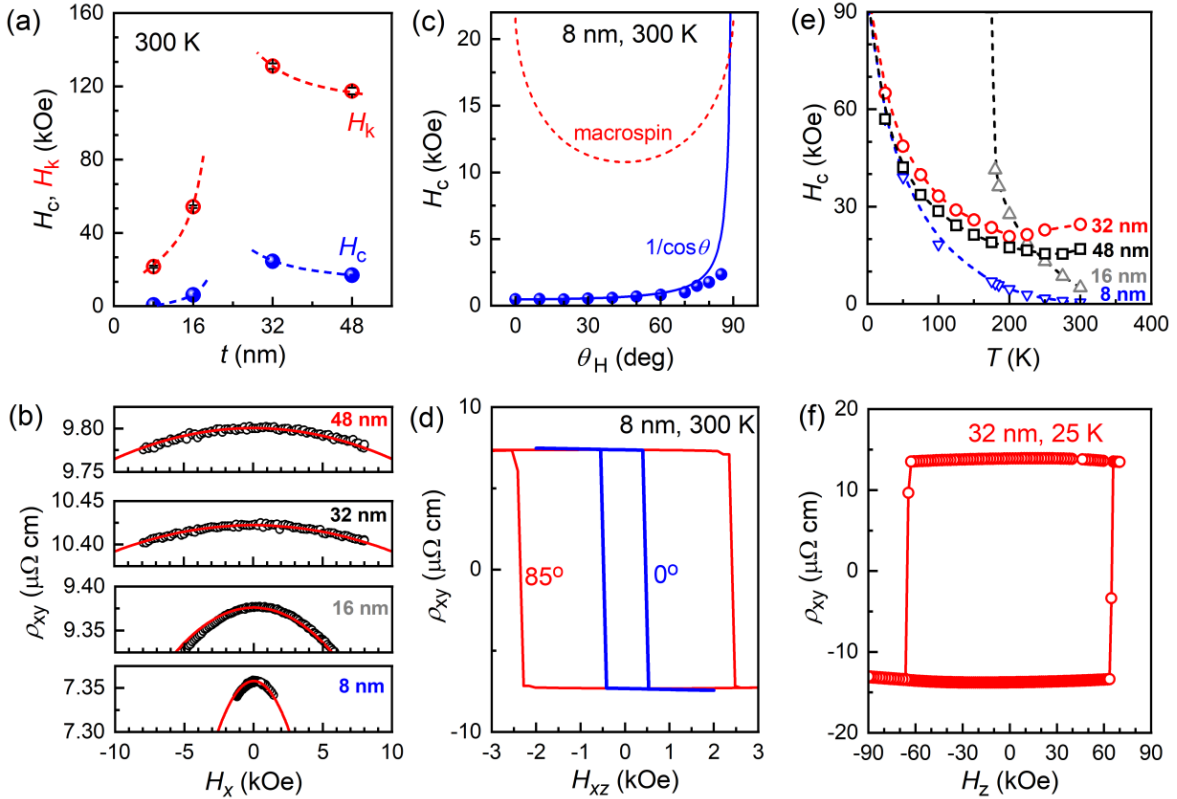

Fig. 3. (a) Perpendicular coercivity and perpendicular magnetic anisotropy field and (b) Parabolic scaling of the transverse resistivity with in-plane magnetic field for FeTb with different thicknesses (300 K). (c) Dependence on the polar angle ($\theta_H$) of the room-temperature coercivity. (d) Transverse resistivity hysteresis loops for the 8 nm FeTb measured at $\theta_H = 0°$ and $85°$ and 300 K. (e) Dependence on the temperature of the perpendicular coercivity ($\theta_H = 90°$) of the FeTb films. (f) Transverse resistivity hysteresis loop at 25 K for the 32 nm FeTb, displaying a giant coercivity of 65 kOe.

## IV. Resistivity upturn

Resistivity or electron momentum scattering is also a key property of a spintronic material. For instance, electron momentum scattering affects spin-dependent scattering, the generation and relaxation of spin current via SOC. To provide insight into the electron momentum scattering mechanism, we measure the resistivity of the FeTb samples as a function of temperature (Fig. 4(a)). $\rho_{xx}$ varies between 210 μΩ cm and 250 μΩ cm. In analogy to the magnetic properties and the anomalous Hall resistivity (see below), $\rho_{xx}$ shows also an interesting non-monotonic thickness dependence at each fixed temperature due to some exotic mechanism yet to know. Here, interfacial scattering is unlikely to play any significant role in the determination of $\rho_{xx}$ of these thick, resistive FeTb because they should have a very short mean-free path.

More interestingly, $\rho_{xx}$ shows a quasi-linear upturn upon cooling for each sample. Similar resistivity upturn has also been observed in 200 nm thick FeTb films [35]. In general, a resistivity upturn can arise from weak localization, hopping conductance, orbital one-channel Kondo effect, orbital two-channel Kondo effect, electron-electron scattering, magnetic Brillouin zone scattering, or scattering of electrons by thermal vibration of structure factor. However, none of these mechanisms appear to explain the resistivity upturn of these FeTb films. First, weak localization, which diminishes under an external magnetic field or a strong internal exchange field, is not expected in the ferrimagnetic FeTb that have giant perpendicular magnetic anisotropy (Fig. 3(a)). Hopping conductance is known to occur in Mott-Anderson insulators with extremely high resistivity (e.g., $10^{6}$-$10^{9}$ μΩ cm for quasicrystal AlPdRe [36], amorphous GeTe, and GeSb$_2$Te$_4$ annealed at 150 °C [37], $10^{11}$-$10^{17}$ μΩ cm for the Pt-SiO$_2$ granular film with Pt concentration of 0.11 [38,39]) but not in metals like FeTb with several orders of magnitude lower resistivity. The absence of hopping conductance is reaffirmed by the lack of a $T^{-1/4}$ scaling (so-called Mott's law [40]) in the resistivity (Fig. 4(b)). The orbital one-channel Kondo effect [41], if important, should increase the resistivity as a function of ln$T$, which is not the case for the FeTb (Fig. 4(c)). The resistivity upturn due to electron-electron interaction would follow a $T^{1/2}$ scaling at low temperatures [42], which is not consistent with the evident deviation from the $T^{1/2}$ scaling at temperatures below 150 K (Fig. 4(d)).

The orbital two-channel Kondo effect [41,43-45] is also less likely to explain the resistivity upturn in the amorphous FeTb films. This is because it would imply a Kondo temperature of >300 K (below which the $T^{1/2}$ scaling emerges) and a deviation temperature of 150 K (below which the resistivity deviates from the $T^{1/2}$ scaling), both of which are surprisingly high. Note that the Kondo temperature is only 23 K for the $L1_0$-MnAl [42] and 14.5 K for $L1_0$-MnGa films [45], a few K for glasslike ThAsSe [43] and Cu point contacts [41], while the deviation temperature is typically below 1 K for all the previously studied two-channel Kondo systems [41-43].

Another possible mechanism for a resistivity upturn is the magnetic Brillouin zone scattering (the periodic potentials due to antiferromagnetic alignment of the magnetic sublattices can produce an additional magnetic Brillouin zone, of smaller volume in $k$-space than the ordinary lattice potential, whose planes further incise and contort the Fermi surface [47]). While this possibility cannot be quantitatively tested due to a lack of knowledge about the exact functional dependence on temperature, magnetic Brillouin zone scattering should be weak in the amorphous FeTb which has no long-range periodicity in the crystalline and magnetic lattices. The resistivity upturn is also absent in epitaxial ferrimagnets of Mn$_{1.5}$Ga [28] and Mn$_2$Ga [48] in which the two Mn sublattices are also AF coupled.

After we have excluded any important role of weak localization, hopping conductance, orbital one-channel Kondo effect, orbital two-channel Kondo effect, electron-electron scattering, and magnetic Brillouin zone scattering, scattering of electrons by thermal vibrations of the structure factor [49,50] is left as the most likely mechanism for the increase of resistivity with decreasing temperature over a wide range of temperature in our amorphous FeTb films. Thermal vibrations of the structure factor have been reported to explain the resistivity upturn in many liquid transition metals and metallic glass alloys [35,49,50]. Note that such thermal vibrations of the structure factor in disordered alloys are distinct from the phonon scattering that increases the resistivity with increasing temperature in ordered crystalline materials [27-29,44]. Future theoretical calculations of the structure factor as a function of temperature would be informative for a more quantitatively understanding of the resistivity of the FeTb samples, which is, however, beyond the scope of this article.

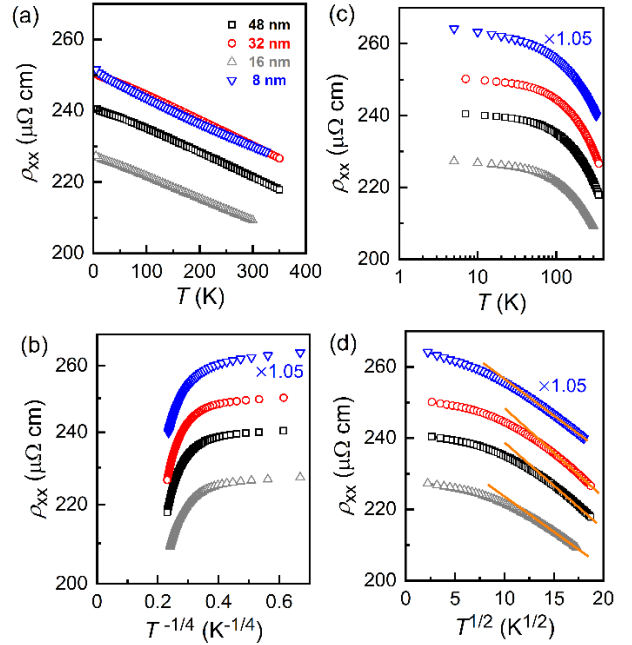

Fig. 4. Resistivities ($\rho_{xx}$) of the FeTb films plotted as a function of (a) temperature $T$, (b) $T$ (in log plot), (c) $T^{-1/4}$, and (d) $T^{1/2}$. In (b)-(d) the resistivity data for the 8 nm FeTb is multiplied by 1.05 for clarity. In (b) the log plot of $\rho_{FeTb}$ as a function of $T^{-1/4}$ indicates a lack of Mott's law for hopping conduction [38], the latter predicts ln$\rho_{xx}$ to vary linearly with $T^{-1/4}$. In (d) the straight lines represent the best linear fits to the data in the high-temperature regime.

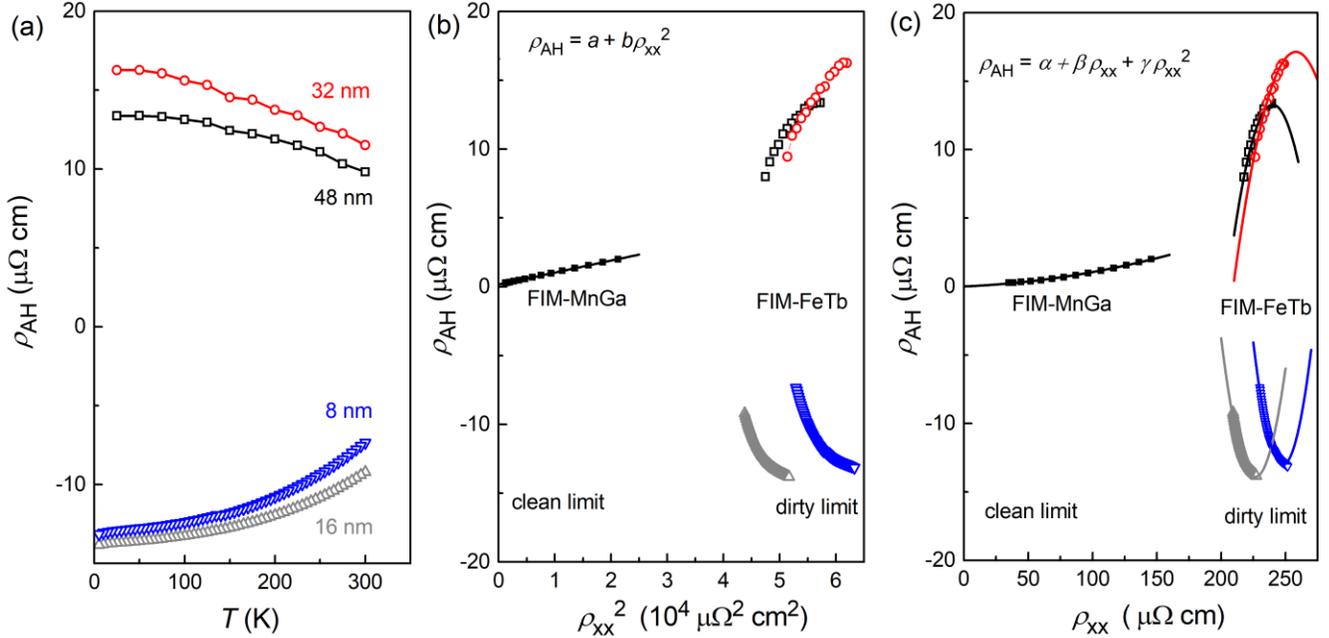

Fig. 5. Scaling of the anomalous Hall effect. (a) Dependence on the temperature of $\rho_{AH}$ of the FeTb films with different thicknesses. (b) $\rho_{AH}$ vs $\rho_{xx}^2$ and (c) $\rho_{AH}$ vs $\rho_{xx}$ for the FeTb films and the control $Mn_{1.5}Ga$ sample. The solid straight lines in (b) represent fits of the data to Eq. (3) and the solid curves in (c) represent the fits of the data to Eq. (4).

## V. Scaling of the strong Anomalous Hall effect

We now discuss the scaling of the anomalous Hall resistivity ($\rho_{AH}$) with the longitudinal resistivity ($\rho_{xx}$). The scaling analysis is interesting as it can disentangle the intrinsic and extrinsic contributions of the anomalous Hall resistivity of magnetic materials in which the electron scattering is dominated by impurity and phonon scattering. In that case, $\rho_{AH}$ of a given sample is simply a linear function of $\rho_{xx}^2$, i.e.,

$$\rho_{AH} = \alpha\rho_{xx0} + \beta_0 \rho_{xx0}^2 + b\rho_{xx}^2, \quad (3)$$

where $\alpha$, $\beta_0$, $\rho_{xx0}$, and $b$ are constant for a given sample and $a = \alpha\rho_{xx0} + \beta_0 \rho_{xx0}^2$ goes to zero when the residual resistivity $\rho_{xx0}$ (due to static impurity scattering at low temperatures) is zero. Equation (3) describes the AHE scaling of epitaxial FIM $Mn_{1.5}Ga$ [28] and some other 3$d$ ferromagnets. Hou *et al*. [29] also proposed a multivariable scaling relation for the AHE in magnetic materials in *very high-conductivity* regime by assuming two major competing scattering sources: i.e.

$$\rho_{AH} = \alpha\rho_{xx0} + \beta_0 \rho_{xx0}^2 + \gamma_0 \rho_{xx0} \rho_{xxT} + \beta_1 \rho_{xxT}^2, \quad (4)$$

where $\rho_{xx0}$ is also the residual resistivity and $\rho_{xxT} = \rho_{xx} - \rho_{xx0}$ is resistivity due to dynamic phonon scattering at high temperatures, $\alpha$, $\beta_0$, $\gamma_0$, and $\beta_1$ are fitting parameters. Equation (4) describes well the AHE scaling in epitaxial ferromagnetic Fe [27] grown by molecular-beam epitaxy. For convenience, we rewrite Equation (4) as

$$\rho_{AH} = \alpha\rho_{xx0} + (\gamma_0 - 2\beta_1)\rho_{xx0}\rho_{xx} + (\beta_0 + \beta_1 - \gamma_0)\rho_{xx0}^2 + \beta_1 \rho_{xx}^2, \quad (5)$$

We note that Eq. (3) -Eq. (5) predict that $\rho_{AH}$ is a monotonic function of $\rho_{xx}$ and scales smoothly to zero at zero $\rho_{xx}$ ($\rho_{xx0} = \rho_{xxT} = 0$).

In Fig. 5(a), we plot the values of $\rho_{AH}$ for the FIM FeTb as a function of temperature. While the anomalous Hall resistivities of the Fe-dominated and Tb-dominated films are of opposite signs, the magnitude increases monotonically, by 50%, for each sample. A similar increase of $\rho_{AH}$ with temperature has also been reported in ferromagnetic MnAl with orbital two-channel Kondo effect [51] and is distinct from that of FMs (e.g., Fe [27], Co [52], Ni [53], and FePt [54]) and FIMs (MnGa [28]) in which the electron scattering is dominated by impurity scattering and phonon scattering. More surprisingly, $\rho_{AH}$ of the FeTb is not a linear function of $\rho_{xx}^2$ and even does not have an obvious monotonic scaling towards zero as $\rho_{xx}$ decreases. This observation suggests the breakdown of the conventional AHE scaling for the dirty metal of amorphous FeTb ferrimagnets. This breakdown is not a general case for FIMs as the $Mn_{1.5}Ga$ with AF-coupled Mn sublattices does follow Eq. (3) (see Fig. 4(b)).

We show in Fig. 5(c) that the anomalous Hall resistivity phenomenally follows the law

$$\rho_{AH} = \alpha + \beta\rho_{xx} + \gamma\rho_{xx}^2, \quad (6)$$

where $\alpha$, $\beta$, and $\gamma$ are non-zero constants. Equation (6) predicts a peak and decay in $\rho_{AH}$ at very high resistivities which might be consistent with the expectation that $\rho_{AH}$ should reduce towards zero in the limit of infinite $\rho_{xx}$ (insulators). Note that we have tested that the data in Fig. 5(c) cannot be fit by a monotonically varying exponential, logarithm, hyperbola, and other functions. The underlying physics and the precise application regime of the new scaling, Eq. (6), require theoretical and experimental investigations in the future and is beyond the scope of this work.

Finally, we mention that the anomalous Hall resistivity of the FeTb is giant compared to that of the 3$d$ magnets Fe [27], Co [52], Ni [53], $Co_{40}Fe_{40}B_{20}$ [55], $Mn_{1.5}Ga$ [28], MnAl [51], and Mn$_3$Ge [56] (Fig. 6(a)) due to the large anomalous Hall angle ($\rho_{AH}/\rho_{xx}$, see Fig. 6(b)) and the high resistivity. As plotted in Fig. 6(c), the anomalous Hall conductivity of the FeTb is also stronger than that of MnAl, MnGa, Mn$_3$Ge, and CoFeB with significantly higher longitudinal conductivities. Such giant anomalous Hall effect is highly preferred for sensor applications.

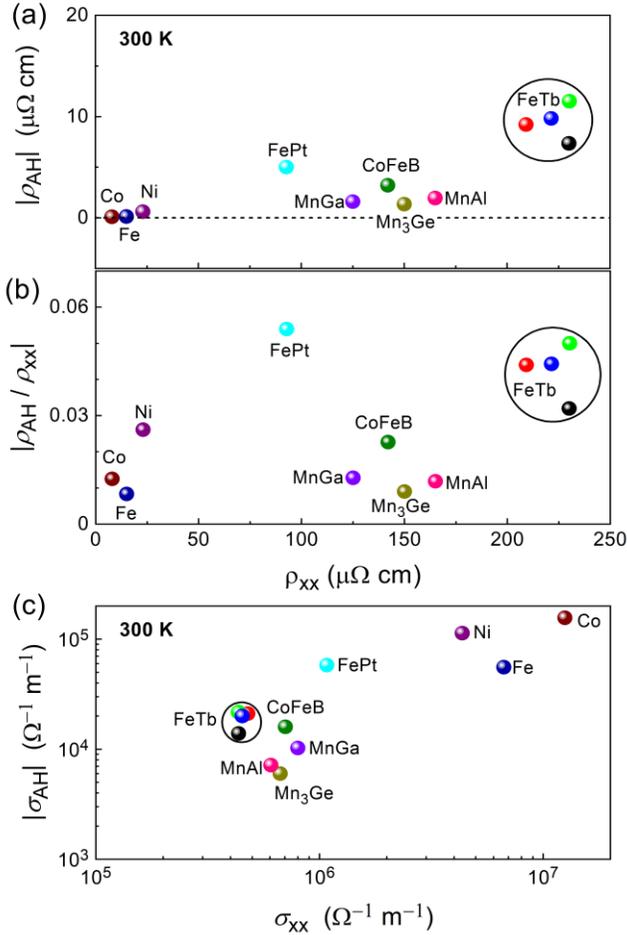

Fig. 6. Dependence on the longitudinal resistivity of (a) the anomalous Hall resistivity and (b) the anomalous Hall angle of representative magnetic films. (c) The anomalous Hall conductivity of the same materials plotted as a function of the longitudinal conductivity.

**Conclusion**

We have presented a systematic study of the magnetic and transport properties of the ferrimagnetic FeTb alloy by varying the layer thickness and temperature. The FeTb is tuned from the Tb-dominated regime to the Fe-dominated regime simply via the increase of the layer thickness, without varying the composition. For each of the studied FeTb samples, the coercivity closely follows the $1/\cos\theta_H$ scaling (where $\theta_H$ is the polar angle of the external magnetic field) and increases quasi-exponentially upon cooling and exceeds 90 kOe below a certain low temperature, revealing that the nature of the coercivity is thermally-assisted domain wall depinning field. The resistivity increases quasi-linearly with temperature upon cooling likely due to thermal fluctuations of the structure factor of the amorphous FeTb. The anomalous Hall resistivities of both Fe- or Tb-dominated FeTb layers that are in the dirty limit cannot be described by any of the existing AHE scaling laws proposed in the literature. These exotic findings should advance the understanding of the magnetic and transport behaviors of the transition-metal-rare-earth ferrimagnets.

The authors thank Changmin Xiong for help with PPMS measurements. This work is supported partly by the National Key Research and Development Program of China (Grant No. 2022YFA1204004), partly by the National Natural Science Foundation of China (Grant No. 12274405), and partly by the Strategic Priority Research Program of the Chinese Academy of Sciences (XDB44000000). The EDS measurements performed at Shaanxi Normal University was supported by the National Natural Science Foundation of China (Grant No. 51901121).